\documentstyle[12pt]{article}

\begin{document} 
\begin{titlepage}

Preprint \hfill   \hbox{\bf SB/F/96-237}

\hfill 	\hbox{\bf CBPF-NF-027/96}

\hfill   \hbox{hep-th/9605074} 
\hrule 
\vskip 1.5cm
\centerline{\bf On Topological Field Theories and Duality}   
\vskip 1cm

\centerline{J.Stephany$^{1,2}$}  
\vskip 4mm
\begin{description}

\item[]{\it  $^1$ Centro Brasilero de Pesquisas
F\'{\i}sicas, Departamento de Part\'{\i}culas e Campos, Rua
Xavier Sigaud 150, Urca, Rio de Janeiro, RJ, Brasil.}
\item[]{\it $^2$ Universidad Sim\'on
Bol\'{\i}var,Departamento de F\'{\i}sica,Apartado Postal
89000,  Caracas 1080-A, Venezuela.}  
\item[]{\it \ \ \ e-mail: stephany@cbpfsu1.cat.cbpf.br}    
\end{description}
\vskip 1cm

{\bf Abstract} \vskip .5cm

Topologically non trivial effects appearing in the
discussion of duality transformations in higher genus
manifolds are discussed in a simple example. Their
relation with the properties of Topological Field Theories
is established.

\vskip 2cm  
\hrule  
\bigskip  
\centerline
{\bf UNIVERSIDAD SIMON BOLIVAR}  
\vfill 
\end{titlepage}

Duality transformations \cite{Dual} \cite{Spin} \cite{2D}
\cite{SW} \cite{FDual} are constructed with the aim to
relate different models of  particles, strings and other
extended objects by  establishing equivalences between their
spectrum and observables.  For spin systems \cite{Spin} and
some two dimensional field theories \cite{2D} they have
been constructed explicitly and present interesting
features like strong coupling to weak coupling mappings or
definite relations between solitons and Fock space states.
However in most of the cases  where their existence have
been conjectured only partial evidence of the desired
correspondences have been established, mainly in the form of
particle spectrum identifications. In the path integral
approach  the search of duality transformations translate
to that of an adequate equivalence between partition functions
or generating functionals. There, space is also opened to
apply related ideas to  restricted low energy effective
actions \cite{SW}. 

Dualized models have been obtained 
using path integrals introducing auxiliary fields in the
path integral conveniently restricted and integrating out
the original fields (or part of them)\cite{FDual}. For 2-D
and some 3-D models the results obtained by this method
\cite{2-3D} have been shown to match those obtained in the
operatorial approach \cite{Spin} \cite{2D}.  In this letter
we show by taking a simple model that this procedure
corresponds to a coupling with a topological field theory.
This introduces the topological properties of the base
manifold into the formalism, and  gives a dynamical
function to the topological field theories.  

There are essentially two forms in which  a
least action principle  implements a
linear restriction of the form $ G^{ij} \varphi _j = 0$:
Introducing a quadratic lagrangian density ${\cal
L}= \varphi_i G^{ij} \varphi _j$ or by means of a Lagrange
multiplier. In the path integral approach   the same 
effect of incorporating a factor of a power of $1/det (G)$ is obtained.
Using the Lagrange multiplier one has as an intermediate
step    \begin{equation}  
\label{delta} 
I(\varphi) = \int {\cal D}\varphi \delta
(G\varphi)exp-\int {\cal L}(\varphi)d^Dx ,   
\end{equation}  which
allows for additional factors.
This situation is somehow modified when the operator $G$
applied to the fields is non-singular as occurs with gauge
systems. In this case, care has to be taken with the zero
eigenvalues of the the operator by means of some procedure
which ultimately corresponds to the introduction of
a modified measure. One has also the additional restriction
of looking only to gauge invariant aspects of the model.
This is  the situation one faces when one tries to impose
the restriction   $F_{\mu \nu}(A)=0$, or more generally,
$dA=0$ on gauge fields. If no other fields are involved
after taking care of the longitudinal sector in the path
integral, the two options above  correspond to
nothing else but a Chern-Simons-like topological
field theory   \cite{CS} or  a topological $BF$  model of coupled
antisymmetric and vector fields \cite{BF},\cite{BF2}. The
correct  integration measure is best
obtained by imposing $BRST$ invariance of the effective
action. This  leads to the complete definition of the
corresponding topological field theories \cite{CS}\cite{BF}
\cite{BF2}. In this sense and stressing the structure of
(\ref{delta}) we note that $BF$ theories are the
adequate tools to define the restrictions $\delta(dA)$ or
$\delta(F_{\mu \nu}(A))$ into the path integral framework.
Going into the details let us write  the partition
function for such models \cite{BF} \cite{BF2}   
\begin{equation}  
\label{BF} 
Z[0]=\int {\cal D}A {\cal D}B
{\cal D}h e^{-\int ({\cal L}_{BF} + {\cal L}_{gf})d^Dx} , 
\end{equation}  
where ${\cal D}h$ stands for the
integration on the complete set of ghost and auxiliary
fields, ${\cal L}_{gf}$ is the gauge fixing term of the
lagrangian density and ${\cal L}_{BF}$ is the $BF$
lagrangian. This is given in the general case by ${\cal
L}_{BF}=B \land F(A)$ with $B$ a $(D-p-2)$-form,  $A$ an 
$p+1$-form and $F=dA$. 
In the particular case of $D=3$ and
$A$ a vector field we have the simple expression
\begin{equation}  
\label{BFL}  
{\cal L}_{BF} = B_\mu
F^\mu(A) = B_\mu \epsilon^{\mu \nu \rho}\partial_\nu
A_\rho   
\end{equation} 
which will be useful later.
Here, $F_\mu(A)=1/2\epsilon_{\mu \nu
\rho}F^{\nu \rho}(A)$ and  the conditions $F_\mu(A)=0$
and $F_{\mu \nu}(A)=0$ are completely equivalent since
$F_0=F_{12}$, $F_1=F_{20}$ and $F_2=-F_{10}$. 

In  recent works \cite{FDual}, the restriction of zero
curvature imposed to an auxiliary gauge field has been used
as a fundamental ingredient for the introduction of dual
variables and  dualized models in the path integral
approach. The essential steps of this method are the
following. First, a  gauge symmetry is identified and
the corresponding gauge model is considered restricted to
the condition $F_{\mu \nu}(A)=0$. This is implemented by
means of a Lagrange multiplier. In  fact, as we show below
in a concrete example  the symmetry considered may be one
of only some terms of the lagrangian density and the main
line of reasoning remains untouched. 
Second, after some intermediate manipulations which depend 
 on the specific model considered, the auxiliary
field or the lagrange multiplier becomes  
the fundamental variable of the dualized
model.    The appearance or not of a mapping between the
strong coupling and the weak coupling of the models is not
granted by this procedure and depends of the systems under
consideration. 

Since topological field theories are distinguished for being
able to extract  the topological non-trivial information of
the manifolds where they are formulated, stressing their
role in the construction of the dualized models appears as a
promising way of incorporating this issues in the
formulation. In what follows  we will show how global
aspects intervene the implementation of duality in the
rather simple but non-trivial example of vector models in
3-D.   

In 3-D massive, parity odd excitations may be described by
three different vector models \cite{VM} which are
respectively the topologically massive model (TMM), the
so-called self-dual model (SDM) (here self-dual is not
related with duality as we are interested but refers to a
property of the equations of motion of the model) and a
third model which we will call the intermediate model (IM).
The corresponding lagrangian densities are given by 
\begin{equation} 
\label{TMM} 
{\cal  L}_{TMM}=-{m \over 2}
(\epsilon^{\mu \nu \rho}\partial_\nu A_\rho)(\epsilon^{\mu
\alpha \beta}\partial_\alpha A_\beta) +  {1 \over 2}A_\mu
\epsilon^{\mu \nu \rho}\partial_\nu A_\rho
\end{equation}   
\begin{equation} 
{\cal L}_{SDM}={m \over 2}a_\mu a^\mu -{1 \over 2}a_\mu
\epsilon^{\mu \nu \rho}\partial_\nu a_\rho
\label{SDM} 
\end{equation}  
\begin{equation} 
\label{IM}
{\cal L}_{IM}={m \over 2}a_\mu a^\mu - a_\mu \epsilon^{\mu
\nu \rho}\partial_\nu A_\rho +  {1 \over 2}A_\mu
\epsilon^{\mu \nu \rho}\partial_\nu A_\rho 
\end{equation} 
These systems have been studied extensively from various
points of view and  may be shown to be locally
equivalent by means of different analysis. Deser and Jackiw
\cite{VM} provided the original proof of the equivalence 
between the (TMM) and the (SDM) solving the canonical 
equal-time
algebra of the quantized fields in terms of a canonical
free massive field. They also introduced the intermediate
model as a master first order formulation of the other two:
Taking variations respect to $a_\mu$ or $A_\mu$ in
(\ref{IM}) and substituting back the resulting equation in
${\cal L}_{IM}$, one recovers respectively ${\cal L}_{TMM}$ or
${\cal L}_{SDM}$. The local equivalence of this models
has also been discussed in the canonical Hamiltonian
approach \cite{GRS} and in fact it has been shown that the
SDM, which is not a gauge theory, emerges as a gauge fixed
version of the TMM in topologically trivial manifolds. On
the other hand  in higher genus manifolds, the TMM and the
SDM are not equivalent. This is most easily established
noting that the only  solution of the SDM which satisfies
$F_\mu (a)=0$ is $a_\mu=0$ in contrast to the TMM for which
every flat connection is a solution \cite{AR}. 

Let us turn to the point we want to raise and note that the
TMM may also be obtained as the dualized version of SDM 
when  one applies the duality transformation described
above. For genus zero manifolds, where the two systems are equivalent this provides just another way to show this  equivalence. For higher genus manifolds  as discussed above the systems are not globally equivalent and so we have a concrete example for which the duality transformation induces non trivial topological properties in the resulting model. To see this let us consider the partition function
of the SDM in a genus zero manifold  
\begin{equation}
\label{ZSD} 
Z_{SDM}[0] = {\cal N}\int {\cal D}a_\mu exp-\int\Bigl(
{m \over 2}a_\mu a^\mu -{1 \over 2}a_\mu \epsilon^{\mu \nu
\rho}\partial_\nu a_\rho) \Bigr)d^3x .  
\end{equation} 
For notational simplicity we write our equations for a locally flat metric but they generalize to the curved case. Next observe that the second term in ${\cal L}_{SDM}$ is invariant under the addition of a gradient. In genus zero manifolds if one introduces an auxiliary gauge field$ A_\mu$
coupled to $a_\mu$ in the form 
\begin{equation}
\label{INT} 
{\cal L}_{int}(a,A)=-{1
\over 2}(a_\mu + A_\mu) \epsilon^{\mu \nu \rho}\partial_\nu
(a_\rho + A_\rho)
\end{equation} 
and impose 
\begin{equation}
\label{ZC}
F_\mu(A)=0,
\end{equation} 
in order to perform the duality transformation, then, 
$A_\mu=\partial_\mu\Lambda$
is a pure gauge  and  ${\cal L}_{int}(a,A)$ is in fact
equal to the second term of ${\cal L}_{SDM}$. 
In higher genus manifolds this  is
not true because there are solutions to (\ref{ZC}) 
which cannot be written globally as pure gauges and is here that the non-trivial
topological properties of the system find their
way into the formulation. After introducing 
in such a way the dual variables we have in an arbitrary manifold,
\begin{eqnarray}
\label{ZDual} 
Z_{SDM}^{Dual}[0] = {\cal N}\int {\cal D}A_\mu {\cal D}a_\mu \delta(F_\mu(A))exp-\int\Bigl({m \over 2}a_\mu a^\mu \nonumber\\ -{1
\over 2}(a_\mu + A_\mu) \epsilon^{\mu \nu \rho}\partial_\nu
(a_\rho + A_\rho) + \hbox{gauge fixing terms}\Bigr)d^3x .  
\end{eqnarray} 
We introduce a Lagrange multiplier $B_\mu$ to
promote the $\delta(F_\mu(A))$ to the lagrangian.
To maintain  our argument simple, we do
not enter into the details of the gauge fixing procedure,
which are well understood \cite{BF} \cite{BF2} and amount 
to a proper definition
of $\delta(F_\mu(A))$ and simply raise to the effective
lagrangian  a gauge fixing term for  each  auxiliary
field. 
To facilitate the Gaussian integration that
follows, we choose the conditions $\partial^\mu(A_\mu +
a_\mu - B_\mu)=0$ for the $A$ field and $\partial^\mu
B_\mu=0$ for the $B$ field which are clearly allowed.
We then have,
 \begin{eqnarray}  
 \label{ZBF}
 Z_{SDM}^{Dual}[0] =
{\cal N}\int {\cal D}a_\mu {\cal D}B_\mu {\cal D}A_\mu  exp
-\int \Bigl(-{1 \over 2}(a_\mu + A_\mu) \epsilon^{\mu \nu
\rho}\partial_\nu (a_\rho + A_\rho) \nonumber\\ + {m \over
2}a_\mu a^\mu  + B_\mu(\epsilon^{\mu \nu \rho}\partial_\nu
A_\rho) + {1\over{2\chi}} (\partial_\mu B^\mu)(\partial_\nu
B^\nu)  \nonumber\\ + {1 \over {2\xi}}\partial^\mu(A_\mu +
a_\mu - B_\mu)\partial^\nu(A_\nu + a_\nu - B_\nu)
\Bigr)d^3x.  
\end{eqnarray}     
 What we
have obtained in this intermediate step is the partition
function of a $BF$ topological field theory coupled to a
matter field described by  the SDM. 
Now, one can perform the regular Gaussian integration in the field  
${\tilde A}_\mu=A_\mu + a_\mu - B_\mu$, and we get
\begin{eqnarray} 
\label{ZIM} 
Z_{SDM}^{Dual}[0] = {\tilde{\cal N}} \int {\cal D}a_\mu 
{\cal D}B_\nu 
exp-\int \Bigl( {m \over 2}a_\mu a^\mu - a_\mu
\epsilon^{\mu \nu \rho}\partial_\nu B_\rho \nonumber\\ +
{1 \over 2}B_\mu \epsilon^{\mu \nu \rho}\partial_\nu
B_\rho + {1\over{2\chi}} (\partial^\mu B_\mu)(\partial^\nu
B_\nu) \Bigr)d^3x = Z_{IM}[0]. 
\end{eqnarray} 
This is the partition function of the IM,
which may be also recognized as a Chern-Simons topological model coupled to the SDM.  By simply performing the
Gaussian integration in $a_\mu$,  we  obtain directly as it was advanced  the
partition function of the TMM. For the reasons mentioned above, the dualized model we end up is not globally equivalent to the one we started with. This can be shown most clearly at this point by factorizing in the partition function the term which encodes the topological information of the manifold. To this end we take advantage of the gauge invariance of the system and proceed in the following way. Instead of integrate $a_\mu$ in (\ref{ZIM}), we can make the shift 
\begin{equation}
B_\mu\rightarrow B_\mu + a_\mu
\end{equation}
This leaves us with
\begin{eqnarray}
\label{ZOM} 
Z_{SDM}^{Dual}[0] = {\tilde{\cal N}} \int {\cal D}a_\mu 
{\cal D}B_\nu 
exp-\int \Bigl( {m \over 2}a_\mu a^\mu - {1 \over 2}a_\mu
\epsilon^{\mu \nu \rho}\partial_\nu a_\rho \nonumber\\ +
{1 \over 2}B_\mu \epsilon^{\mu \nu \rho}\partial_\nu
B_\rho + {1\over{2\chi}} \partial^\mu (B_\mu+a_\mu)\partial^\nu
(B_\nu+a_\nu) \Bigr)d^3x . 
\end{eqnarray}
After recognizing that $\partial^\mu(B_\mu+a_\mu)=0$ is an acceptable gauge fixing condition for the gauge field $B_\mu$ we end up with the factorized relation
\begin{equation}
Z_{SDM}^{Dual}=Z_{CS}Z_{SDM}
\end{equation}
where $Z_{CS}$ is the partition function of the Chern Simons topological theory. This is confirmed by a detailed computation in the Hamiltonian approach\cite{AR}. Some
of this considerations generalize also to non-Abelian 
and tensor fields  \cite{ALR} \cite{AS}. 

A similar relation between the generating functionals of
the  models may also be obtained such
that if we introduce the external current minimally coupled
to the TMM (which is a gauge model and calls for it) we do
not get this current minimally coupled to the SDM. This,
together with the topological blindness of the SDM is
relevant for the discussion of anyons within these models.
\cite{anyonic}.

Let us conclude by summarizing the most salient lessons we
take from this analysis: 
\begin{itemize} 
\item{} Duality transformations are implemented by  coupling
 the original model with a $BF$ topological theory. In
genus zero manifolds this do not introduce any difference
but in higher genus manifolds the global equivalence of
the models isno longer valid. In the case discussed above
the net effect is a coupling of the
matter fields with a Chern-Simons topological theory. This
feature is likely to be generalized to other contexts and
furnishes a dynamical function for the topological fields,
theories as mentioned  at the beginning. We note that
this matches with the fact that although $BF$
fields interacting with classical sources  do not 
act with  a force on them, they select the
allowable trajectories on topological grounds \cite{LRS}.
\item{}Our experience with the TMM and the
SDM suggests also to look to models connected by a duality
transformation as related by a gauge fixing procedure
\cite{GRS}. We note that the physical observables in the TMM
are only the gauge invariant operators and this does not
occur in the SDM for which other operators are also allowed
as observables.   
\end{itemize}

The issues discussed in this letter do not address the
interesting possibility of duality  between the
particles of the TMM and the SDM and the soliton spectrum
of this or related models. On the other hand, most of the
discussion presented here translate to more general contexts
where duality transformations have been implemented in the
functional approach. The conclusions derived from this 
minimal model should shed light to these more general cases.
In particular one understands in a simple way why
the dualized models should
become sensible to the topological properties of the base
manifold. \vskip .3cm     

I wish to thank the Centro Latinoamericano de F\`{\i}sica (CLAF)
and the Conselho Nacional de Desenvolvimento Cient\'{\i}fico e
Tecnol\'ogico (CNPq) for financial support and P.J.Arias,
H.R.Christiansen, S.A.Dias, R.Gianvittorio, M.A.Kneipp and
A.Restuccia for useful discussions.

\end{document}